\documentclass[pra,tightenlines,superscriptaddress,12pt]{revtex4}

\usepackage{bm}

\usepackage{graphicx}

\begin{document}

\def\ket#1{|#1\rangle}
\def\bra#1{\langle#1|}
\def\tr{{\rm tr}}

\def\BFM{\mbox{BFM}}

\title{Conditions for compatibility of quantum state assignments}

\author{Carlton M.~Caves}
\email{caves@info.phys.unm.edu}
\affiliation{Department of Physics
and Astronomy, University of New Mexico, Albuquerque, NM~87131-1156,
USA}
\affiliation{Department of Physics and Special Research Centre for Quantum
  Computer Technology, The University of Queensland, Queensland 4072,
  Australia}
\author{Christopher A.~Fuchs}
\email{cafuchs@research.bell-labs.com}
\affiliation{Bell Labs, Lucent Technologies, 600--700 Mountain Avenue, Murray Hill, NJ 07974, USA}
\affiliation{Department of Physics and Special Research Centre for Quantum
  Computer Technology, The University of Queensland, Queensland 4072,
  Australia}
\author{R\"udiger Schack}
\email{r.schack@rhul.ac.uk}
\affiliation{Department of Mathematics, Royal Holloway, University of
London, Egham, Surrey TW20$\;$0EX, UK}
\affiliation{Department of Physics and Special Research Centre for Quantum
  Computer Technology, The University of Queensland, Queensland 4072,
  Australia}

\date{June 13, 2002}

\begin{abstract}
  Suppose $N$ parties describe the state of a quantum system by $N$
  possibly different density operators.  These $N$ state assignments
  represent the beliefs of the parties about the system.  We examine
  conditions for determining whether the $N$ state assignments are
  compatible.  We distinguish two kinds of procedures for assessing
  compatibility, the first based on the compatibility of the {\it
  prior\/} beliefs on which the $N$ state assignments are based and the
  second based on the compatibility of {\it predictive\/} measurement
  probabilities they define.  The first procedure leads
  to a compatibility criterion proposed by Brun, Finkelstein, and
  Mermin [BFM, Phys.\ Rev.~A {\bf 65}, 032315 (2002)].  The second
  procedure leads to a hierarchy of measurement-based compatibility
  criteria which is fundamentally different from the corresponding
  classical situation.  Quantum mechanically none of the
  measurement-based compatibility criteria is equivalent to the BFM
  criterion.
\end{abstract}


\maketitle

\section{Introduction}

There are good reasons \cite{Caves2002a} to view a quantum state not
as representing a true state of affairs, but as a state of knowledge
or, more provocatively, a {\it state of belief}.  This view
corresponds to the Bayesian approach to probability theory, according
to which probabilities are an agent's necessarily subjective {\it
degrees of belief\/} about a set of alternatives.  Different
scientists can have different beliefs about the same physical system,
resulting in different quantum state assignments.  This can arise for
a variety of reasons.  For instance, one of the scientists might have
no access or only partial access to another's measurement results. In
general, $N$ scientists, or {\it parties}, can assign $N$ different
states, pure or mixed, to a given system.

In this paper we are not concerned with how to justify a particular
state assignment.  Instead, we start from a given set of $N$ states,
representing the beliefs of $N$ parties, and ask for conditions for
determining that the $N$ quantum states are {\em compatible\/} (or,
conversely, that they are contradictory).  The conditions we derive
can be viewed as criteria for the mutual compatibility of $N$ quantum
state assignments.

There are two distinct procedures for assessing the compatibility of
quantum states (or, conversely, for uncovering contradictions among
the states).  The first procedure examines the firm beliefs on which
the parties base their state assignments and asks whether these
beliefs are compatible.  This procedure leads to the compatibility
criterion proposed by Brun, Finkelstein, and Mermin (BFM)
\cite{Brun2002a}. The second procedure examines measurement
probabilities predicted by the parties' state assignments and asks
whether these probabilities are compatible.  We show that this
second procedure leads to a hierarchy of measurement-based
compatibility criteria.  The two-party version of one of our
criteria is equivalent to a compatibility criterion proposed by
Peierls \cite{Peierls1991a,Mermin2001a}; thus this criterion
provides an $N$-party generalization of Peierls's criterion, which
we call {\em post-Peierls\/} (PP) compatibility.

The BFM compatibility criterion is based on the compatibility of the
beliefs of the $N$ parties.  We show that the BFM criterion is not
equivalent to any of the measurement-based compatibility criteria.
This means that there is generally no way to use measurements to
confirm the compatibility of states that are BFM compatible or to
reveal the incompatibility of states that are BFM incompatible.

The compatibility criteria we derive can be specialized to classical
probabilities by considering density operators all of which are
diagonal in a common eigenbasis---i.e., commuting density
operators---and by restricting the allowed measurement operators to
be diagonal in the same basis.  There are interesting differences
between the hierarchy of measurement-based criteria in the classical
and quantum cases. Moreover, the classical version of BFM
compatibility is equivalent to the classical version of PP
compatibility, in contrast to the quantum case.

The paper is organized as follows.  In Sec.~\ref{sec:con}, we review
briefly the concepts of Dutch-book consistency and strong Dutch-book
consistency, which provide the foundations for Bayesian probability
theory, and define the notions of contradictory beliefs and
contradictory probability assignments.  In Sec.~\ref{sec:bfm}, we
derive the BFM compatibility criterion for the quantum states of $N$
parties from the requirement that there be a state assignment that
does not contradict the belief of any party.  Section~\ref{sec:meas}
introduces our hierarchy of measurement-based compatibility criteria
and highlights the surprising differences between the classical and
quantum cases.  We consider how the measurement-based compatibility
criteria change when one generalizes from measurements described by
one-dimensional orthogonal projectors to POVMs.  Section~\ref{sec:pp}
focuses on PP compatibility, the only one of the compatibility
criteria for which---as far as we can tell---it is not possible to
formulate a simple universal mathematical condition applicable to
all sets of quantum states in all Hilbert-space dimensions; we
consider nontrivial examples of PP compatibility for three states of
a three-dimensional system. Sec.~\ref{sec:disc} closes the paper with
a summary and discussion.

\section{Consistent probabilities and contradictory beliefs}
\label{sec:con}

The notion of consistent beliefs has its roots in the approach to
Bayesian or subjective probability via consistent betting behavior.
Bayesian probabilities quantify one's degree of belief in the various
alternatives from among a set of possibilities \cite{Bernardo1994a}.
Bayesian probabilities are given an operational definition in terms
of betting behavior.  Suppose that $A$ is willing to place a bet at
odds of $(1-p)/p$ to 1 on the occurrence of some event.  This means
that $A$ is willing to pay in a {\it stake\/} $px$, with the promise
of receiving a {\it payoff\/} $x$ if the event occurs and nothing
otherwise. That $A$ considers this to be a {\it fair\/} bet---she is
willing to accept the bet at these odds no matter what the payoff,
positive or negative---{\it defines\/} $p$ to be the probability that
$A$ assigns to the event.

Suppose now that $A$ makes probability assignments to a set of
events.  We say that $A$'s probability assignments are {\it
consistent\/} (or Dutch-book consistent) if there exists no set of
bets which she regards as fair, but in which she loses for every
outcome that she believes to be possible.  Notice that this is a
purely internal consistency criterion; it refers only to $A$'s
subjective beliefs.

The so-called Dutch-book argument \cite{DeFinettiBook,Earman1992a}
shows that consistency alone implies that $A$'s probability
assignment must satisfy the usual probability axioms: (i)~$p\ge0$,
(ii)~$p(E)=1$ if $A$ believes that event $E$ is certain to occur,
(iii)~$p(E\vee F)=p(E)+p(F)$ if $E$ and $F$ are mutually exclusive,
and (iv)~$p(E\wedge F)=p(E|F)p(F)$ (Bayes's rule).  In Appendix~A we
first review the formulation of Bayesian probabilities in terms of
betting behavior and then give the Dutch-book derivation of the
standard probability rules.

Consistency enforces the probability axioms, but it does not dictate
particular probability assignments, leaving these to whatever way $A$
chooses to translate what she knows or believes into probabilities.
The only exception is in the case of certainty, where consistency
requires that all probabilities be 0 or 1.  Indeed, a consequence of
the probability axioms---it also follows directly from the Dutch-book
argument---is that $p=0$ for any outcome that $A$ believes to be
impossible.  We call the belief that an outcome is impossible a {\it
firm belief}.

Surprisingly, consistency does not imply that $p>0$ for any outcome
that $A$ believes to be possible.  In other words, if $A$ assigns
probability zero to an outcome, one cannot infer from consistency
alone that $A$ believes the event to be impossible.  To make this
inference, we need a slightly stronger version of consistency: we say
that $A$'s probability assignment is {\it strongly\/} (Dutch-book)
{\it consistent\/} \cite{Shimony1955a,Kemeny1955a} if there exists no
bet that she regards as fair in which there is at least one losing
outcome but no winning outcomes among those outcomes she deems
possible.  If $A$'s probability assignment is strongly consistent,
then an outcome has zero probability if and only if she believes it
to be impossible (see Appendix~A).

Dutch-book consistency has to do with the beliefs and probability
assignments of a single party.  Our concern in this paper is
compatibility among many parties.  We say that the beliefs of $N$
parties about a set of alternatives are {\em compatible\/} if there
is at least one alternative that all parties judge to be possible.
Conversely, the beliefs of $N$ parties are {\em contradictory\/} (or
incompatible) if they are not compatible in the sense just described,
i.e., if every alternative is deemed impossible by at least one
party.  Examining the alternatives to determine which applies is thus
guaranteed to contradict at least one party.  Contradictory beliefs
cannot be reconciled unless at least one party abandons a firm prior
belief.

It is easy to imagine situations that give rise to compatible or to
contradictory beliefs.  Take a die as an example.  Suppose $A$ has
seen the North face of the die, which shows 4 dots.  She therefore
believes that it is impossible for the top face of the die to show
either 3 or 4 dots, but that it is possible for it to show 1, 2, 5,
or 6 dots.  Suppose $B$ has seen the East face of the die, which
shows 1 dot.  He believes that the top face can show 2, 3, 4, or 5
dots, but not 1 or 6.  These beliefs are compatible since both
parties believe that the top face can show 2 or 5 dots.

Now suppose that $C$ asserts that the South face of the die shows 5
dots.  He believes that it is impossible for the top face of the die
to show 2 or 5 dots, thus contradicting the beliefs of $A$ and $B$.
This situation could arise if $C$'s assertion was based on a mistaken
observation from across the room; the beliefs of the three parties
could be reconciled by $C$'s observing the South face again, finding
that it shows 3 dots, and thereby giving up his firm belief that the
top face could not show 2 or 5 dots.  Another possibility is that the
die was tossed after the initial observations by $A$ and $B$; the
beliefs of the three parties could be reconciled if $A$ and $B$
realized that the die had been tossed after their observations, which
would cause them to abandon their prior firm beliefs.

Under the assumption of strong consistency, where the firm belief
that an alternative is impossible is equivalent to assigning zero
probability, the conditions for contradictory and compatible beliefs
can be re-expressed in terms of probabilities.  The beliefs (or
probability assignments) of $N$ parties are compatible if and only if
there is at least one alternative to which all parties assign nonzero
probability, i.e., there exists a probability assignment that does
not contradict the firm beliefs of any of the parties.  This is the
classical version of BFM compatibility.  The beliefs (or probability
assignments) of the $N$ parties are contradictory (or incompatible)
if and only if every alternative is assigned zero probability by at
least one party.

For ordinary consistency, the existence of one or more alternatives
to which all parties assign nonzero probability is sufficient for
compatible beliefs, but it is not necessary, because a party can
assign zero probability to alternatives the party believes are
possible.  For ordinary consistency, probabilities do not carry
enough information about firm beliefs to allow compatibility to be
determined from the parties' probability assignments.  Since we are
interested in the compatibility of density operators, we need strong
consistency so that probabilities for measurement outcomes generated
by the density operators allow one to determine the firm beliefs of
the parties. Therefore we assume strong consistency throughout the
remainder of the paper, except where explicitly noted.

There is a another stronger kind of compatibility for classical
probabilities.  Suppose, for example, that in the case of the die,
$A$ and $B$ come together, combine their observations, and thereafter
agree that the top face can show 2 or 5 dots, but not 1, 3, 4, or 6
dots.  They generally would not assign the same probabilities to 2
and 5 dots, but they do have the same firm beliefs, a situation we
capture by saying that their new beliefs are in concord.  Generally,
we say that the beliefs (or probability assignments) of $N$ parties
are {\em concordant\/} if their firm beliefs coincide, i.e., if they
assign zero probability to the same alternatives.  Concordant
probability assignments have the same support.  It is not reasonable
to demand that parties have concordant probabilities, but they arise
naturally when parties with compatible beliefs share those beliefs.

With this background, we turn now to BFM compatibility for quantum
state assignments.

\section{BFM compatibility}
\label{sec:bfm}

The system under consideration is described by a $D$-dimensional
Hilbert space ${\cal H}$.  We label the $N$ parties by an index
$\alpha=A,B,C,\ldots\;$; their state assignments are denoted by
$\hat\rho_\alpha$.  For a projective measurement in an arbitrary
orthonormal basis $\{\ket k, k=1,\ldots,D\}$, i.e., a measurement
described by {\em orthogonal one-dimensional projectors\/} (we call
such a measurement an ODOP for short), the probability assigned by
party $\alpha$ to the outcome $k$ is given by
$p_k^{(\alpha)}=\tr(\hat\rho_\alpha\hat\Pi_k)=\bra
k\hat\rho_\alpha\ket k$, where $\hat\Pi_k=\ket k\bra k$.  The case of
classical probabilities is included automatically as the situation in
which all the $\hat\rho_\alpha$ are diagonal in the same orthonormal
basis $\{\ket k\}$, and the only allowed measurement is a measurement
in this basis.

Under the assumption of strong consistency, each party assigns zero
probability to precisely those outcomes he believes cannot occur;
i.e., for each $\alpha$, $\bra\psi\hat\rho_\alpha\ket\psi=0$ if and
only if party $\alpha$ believes that the outcome corresponding to
$\ket\psi$ is impossible in any measurement containing $\ket\psi$.
Therefore, each party assigns a density operator $\hat\rho_\alpha$
whose null subspace ${\cal N}(\hat\rho_\alpha)$ consists of all those
vectors corresponding to outcomes he believes cannot occur.  The
support of a density operator is the orthocomplement of the null
subspace. All vectors not in the null subspace have a component in
the support and thus have nonzero probability, so the party believes
that the outcomes corresponding to all such vectors can occur.

A density operator $\hat\rho$ contradicts party~$\alpha$'s firm
beliefs if $\bra\psi\hat\rho\ket\psi>0$ for some $|\psi\rangle$ that
$\alpha$ believes to be impossible.  Thus $\hat\rho$ does not
contradict $\alpha$'s beliefs if and only if ${\cal
N}(\hat\rho_\alpha)\subseteq{\cal N}(\hat\rho)$.  What we want to
know is the circumstances under which there is a density operator
that does not contradict the firm beliefs of any of the parties,
i.e., a density operator $\hat\rho$ such that ${\cal
N}(\hat\rho_\alpha)\subseteq{\cal N}(\hat\rho)$ for all $\alpha$.
Since ${\cal N}(\hat\rho)$ is a subspace, it follows that ${\cal
M}\equiv{\rm span}({\cal N}(\hat\rho_A),{\cal
N}(\hat\rho_B),\ldots)\subseteq{\cal N}(\hat\rho)\subset{\cal H}$.
Such a $\hat\rho$ exists if and only if ${\cal M}$ is not the
entirety of ${\cal H}$, which is equivalent to saying that the
orthocomplement of ${\cal M}$ contains at least one nonzero vector.
Since the orthocomplement of ${\cal M}$ is the intersection of the
supports of the $\hat\rho_\alpha$, we have the result that there
exists a $\hat\rho$ that does not contradict any party's prior belief
if and only if the intersection of the supports of all the
$\hat\rho_\alpha$ contains at least one nonzero vector.  This is the
criterion for {\em BFM compatibility\/} of state assignments.  In the
classical case the BFM criterion reduces to the condition that at
least one of the common eigenvectors has nonzero eigenvalue for all
parties.

What we have shown is that BFM compatibility is equivalent to the
existence of a density operator that does not contradict the firm
beliefs of any party.  Just as in the classical case, the assumption
of strong consistency, as opposed to ordinary Dutch-book consistency,
is essential for this conclusion.  The reason is that {\it any\/} set
of consistent (but not necessarily strongly consistent) state
assignments can arise from a set of noncontradictory beliefs.  Let
$\hat\rho_A,\hat\rho_B,\ldots$ be $N$ arbitrary states.  These are
consistent state assignments for $N$ parties all of whom believe that
{\it any\/} outcome is possible, since consistency alone allows a
party to believe that a vector in his null subspace corresponds to a
possible outcome.  Obviously, there is a posterior state $\hat\rho$
that does not contradict the firm beliefs of any party; indeed, any
posterior state $\hat\rho$ will do.  Merely consistent state
assignments do not reveal enough about the parties' prior beliefs to
rule out the existence of a noncontradictory posterior state
assignment.

Suppose parties with BFM compatible state assignments share their
beliefs, each adopting the firm beliefs of all the others.  BFM
compatibility guarantees that there are density operators that are
consistent with the firm beliefs of all the parties.  The parties
will generally not end up assigning the same density operator, but
they will assign density operators that incorporate the same firm
beliefs and thus have the same support.  We say such density
operators are {\em concordant\/} in the same sense as for probability
assignments.

Our derivation of the BFM criterion is different from the one given
by Brun, Finkelstein, and Mermin \cite{Brun2002a}.  They show that
their criterion follows if one assumes that each of the state
assignments $\hat\rho_A,\hat\rho_B,\ldots$ ``incorporates some subset
of a valid body of currently relevant information about the system,
all of which could, in principle, be known by a particularly
well-informed Zeno.'' Their formulation suggests that each of the $N$
state assignments should be consistent with some real state of
affairs captured in Zeno's state $\hat\rho$.  This impression is
reinforced shortly thereafter in their paper, where one of the
explicit assumptions leading to the BFM criterion is that ``if
anybody describes a system with a density matrix $\hat\rho$, then
nobody can find (the system) to be in a pure state in the null space
of $\hat\rho$.''  In contrast, our derivation is couched wholly in
terms of the beliefs of the parties and does not appeal to a real
state of affairs.  It is therefore preferable in a Bayesian approach
to quantum mechanics.

\section{Measurement-based compatibility}
\label{sec:meas}

\subsection{Compatibility conditions}

We turn our attention now to compatibility conditions based on the
compatibility of measurement probabilities.  We focus first on ODOP
measurements, i.e., those described by complete sets of
one-dimensional orthogonal projectors $\{\hat\Pi_k, k=1,\ldots,D\}$,
the probability party $\alpha$ assigns to outcome~$k$ being
$p_k^{(\alpha)}=\tr(\hat\rho_\alpha\hat\Pi_k)=\bra
k\hat\rho_\alpha\ket k$.  Most importantly, we assume that all
parties agree on this description of the measurement.  In
Sec.~\ref{sec:genmeas} below we generalize the compatibility
conditions to the context of measurements described by POVMs.

Our hierarchy of measurement-based compatibility conditions can be
stated very simply as whether the parties' measurement probabilities
are compatible or concordant and whether this holds for all
measurements or for at least one measurement.  In mathematical
language the compatibility conditions are the following:
\begin{eqnarray}
&&  \forall\{\hat\Pi_j\}\forall k
    \Bigl( ( \forall\alpha:p_k^{(\alpha)}>0 ) \vee
    (\forall\alpha:p_k^{(\alpha)}=0 ) \Bigr)
    \qquad\mbox{(ES)}\;,
\label{allall}   \\
&&  \forall\{\hat\Pi_j\}\exists k \; \forall\alpha:p_k^{(\alpha)}>0
    \phantom{\Bigl(\,((\vee\forall\alpha:p_k^{(\alpha)}=0)) \Bigr)}
    \qquad\mbox{(PP)}\;,
\label{allex}\\
&&  \exists\{\hat\Pi_j\}\forall k
    \Bigl( (\forall\alpha:p_k^{(\alpha)}>0) \vee
    (\forall\alpha:p_k^{(\alpha)}=0) \Bigr)
    \qquad\mbox{(W)}\;,
\label{exall}  \\
&&  \exists\{\hat\Pi_j\}\exists k \; \forall\alpha:p_k^{(\alpha)}>0
    \phantom{\Bigl(\,((\vee\forall\alpha:p_k^{(\alpha)}=0)) \Bigr)}
    \qquad\mbox{(W$'$)}\;.
\label{exex}
\end{eqnarray}

Condition~(\ref{allall}), called ES for ``equal support,'' says that
for all measurements and any outcome $\hat\Pi_k$, either all parties
assign nonzero probability to $\hat\Pi_k$, or they all assign zero
probability to $\hat\Pi_k$.  In other words, the parties'
probabilities for all measurements are concordant.   It is trivial to
see that ES is equivalent to all the density operators
$\hat\rho_\alpha$ having the same support, i.e., being concordant as
defined in the preceding section.  As a consequence, ES implies \BFM.
ES is a very strong compatibility condition, which is violated in
many practical situations, but which arises naturally when parties
with BFM compatible beliefs combine their beliefs.

Unlike ES and BFM, there are fundamental differences between the
classical and quantum versions of the remaining three conditions.
Condition~(\ref{allex}), called PP for ``post-Peierls,'' is implied
by ES.  It says that for all measurements, there is at least one
outcome to which all parties assign nonzero probability; i.e., all
measurements have compatible probabilities.  It is often useful to
think in terms of the conditions for violating PP compatibility: PP
is violated if there exists a measurement such that at least one
party assigns zero probability to every outcome; for this measurement
the measurement probabilities are contradictory, the outcome,
whatever it is, guaranteed to contradict one or more parties.  The
two-party version of PP is the original compatibility condition of
Peierls \cite{Peierls1991a,Mermin2001a}; it is equivalent to
$\hat\rho_A\hat\rho_B\ne0$.  As far as we can tell, the conditions
for multi-party PP compatibility cannot be put in a simple universal
mathematical form, unlike the other compatibility criteria.  In
Sec.~\ref{sec:pp}, we consider nontrivial examples of three-party PP
compatibility in three dimensions.

Condition~(\ref{exall}), called W for ``weak,'' says that there is
at least one measurement such that for any outcome $\hat\Pi_k$,
either all parties assign nonzero probability to $\hat\Pi_k$, or
they all assign probability zero to $\hat\Pi_k$. In other words,
there exists a measurement whose measurement probabilities are
concordant.

Finally, condition~(\ref{exex}), called W$'$, is implied by W. It
states that there is at least one measurement and at least one
outcome for that measurement to which all parties assign nonzero
probability; i.e., there exists a measurement whose measurement
probabilities are compatible.

Summarizing the implications we have identified up till now, we have
\begin{equation}
\label{eq:relall}
\mbox{ES} \Longrightarrow \mbox{BFM} \;, \qquad \mbox{ES}
\Longrightarrow \mbox{PP} \;, \qquad \mbox{W} \Longrightarrow
\mbox{W$'$} \;.
\end{equation}
The latter two relations involve only the logical structure of the
compatibility conditions; all the relations hold for both the
classical and quantum cases.

In the case of classical probabilities, there is only one allowed
measurement---a measurement in the basis that diagonalizes all the
$\hat\rho_\alpha$---so it is clear that ES is equivalent to W and PP
is equivalent to W$'$.  It is equally clear that PP and W$'$ are
equivalent to \BFM.  Summarizing the implications for the classical
case, we have
\begin{equation}
\label{eq:relclassical}
\mbox{Classical probabilities:}\qquad \mbox{W} \Longleftrightarrow
\mbox{ES} \qquad \Longrightarrow \qquad \mbox{BFM}
\Longleftrightarrow \mbox{PP} \Longleftrightarrow \mbox{W$'$} \;.
\end{equation}
This chain reflects the two kinds of compatibility for classical
probabilities: W and ES correspond to the parties having concordant
probabilities, whereas BFM, PP, and W$'$ correspond to the parties
having compatible probabilities.

\subsection{Relations among quantum compatibility conditions}

In the quantum case, the relations (\ref{eq:relclassical}) change in
an interesting way to
\begin{equation}   \label{eq:relquantum}
\mbox{Quantum states:}\qquad \mbox{ES} \Longrightarrow \mbox{BFM}
\Longrightarrow \mbox{PP} \Longrightarrow  \mbox{W$'$}
\Longleftrightarrow \mbox{W}   \;.
\end{equation}
Here, unlike the classical case, BFM is stronger than PP, and PP is
stronger than W$'$, but the most striking difference is that W, the
strongest condition classically, is the weakest condition quantum
mechanically.  As a matter of fact, W is satisfied by {\it any\/} set
of state assignments, as we show below.  The different structure of
quantum implications in Eq.~(\ref{eq:relquantum}) is due to the far
greater freedom quantum mechanics allows for measurements.

We now prove the relations (\ref{eq:relquantum}). The first
implication, $\mbox{ES}\Rightarrow\mbox{BFM}$, is trivial: equal
support implies that the supports have at least one state in common.
It is also clear that the reverse implication does not hold.

To see that BFM implies PP, consider an arbitrary measurement
$\{\hat\Pi_j=\ket{j}\bra{j}\}$.  BFM compatibility is equivalent to
saying that ${\cal M}\equiv{\rm span}({\cal N}(\hat\rho_A),{\cal
N}(\hat\rho_B),\ldots)$ is not the entire Hilbert space ${\cal H}$.
Since the vectors $\{\ket{j}\}$ are an orthonormal basis, at least
one outcome~$\ket{k}$ lies outside ${\cal M}$ and thus has a nonzero
projection onto the orthocomplement of ${\cal M}$, which is the
intersection of the supports of the $\hat\rho_\alpha$.  For this
outcome we have
$\bra{k}\hat\rho_\alpha\ket{k}=\tr(\hat\rho_\alpha\hat\Pi_{k})>0$ for
all $\alpha$.  To see that PP does not imply BFM, consider two
nonorthogonal pure states.  There is no one-shot measurement that can
distinguish the two states reliably, so the two states satisfy PP,
but since the intersection of the supports of the two states contains
only the zero vector, the two states violate BFM.

A simple example shows that W$'$ does not imply PP: two orthogonal
pure states violate PP, but they satisfy W$'$, as can be seen by
considering a measurement in a basis that includes a vector that
lies in the two-dimensional subspace spanned by the two states, but
is not equal to either of them.

To show that W$'$ is equivalent to W and is implied by PP, we prove
the stronger result that {\it any\/} set of $N$ states
$\hat\rho_A,\hat\rho_B,\ldots$ satisfies W, which shows that W
follows from any of the other conditions.  We construct a measurement
$\{\hat\Pi_j\}$ each of whose projectors has nonzero overlap with the
supports of all the $\hat\rho_\alpha$.  Let $\ket{\phi_\alpha}$ be an
eigenvector of $\hat\rho_\alpha$ with nonzero eigenvalue
$\lambda_\alpha$.  We need to find an orthonormal basis $\{\ket{k}\}$
such that $0<|\langle k\ket{\phi_\alpha}|^2$ for all $k$ and
$\alpha$, since this implies $0 < \lambda_\alpha|\langle
k\ket{\phi_\alpha}|^2\le\langle k|\hat\rho_\alpha\ket
k=\tr(\hat\Pi_k\hat\rho_\alpha)$ for all $k$ and $\alpha$, where
$\hat\Pi_k=\ket{k}\bra{k}$.  Letting $S$ be the set of all state
vectors that are orthogonal to at least one $\ket{\phi_\alpha}$, we
see that what we need to do is to construct an orthonormal basis none
of whose basis vectors is in $S$.

To do the construction, we begin by defining the distance between two
state vectors,
\begin{equation}
d(\ket\psi,\ket\chi)\equiv\cos^{-1} |\langle\psi\ket\chi|\;,
\end{equation}
which allows us to define the distance between an arbitrary state
vector $\ket\psi$ and the set $S$ by
\begin{equation}
d(\ket\psi,S)\equiv\min_{\ket\chi\in S} d(\ket\psi,\ket\chi) \;.
\end{equation}
A state vector $\ket\psi$ is in $S$ if and only if $d(\ket\psi,S)=0$.
Our construction relies on the fact that, arbitrarily close to any
vector $\ket\chi\in S$, there exists a vector $\ket\psi$ that is a
finite distance away from $S$; i.e., any $\epsilon$-ball around
$\ket\chi\in S$ contains a vector $\ket\psi$ such that
$d(\ket\psi,S)>0$.

Now choose an orthonormal basis $\{\ket k\}$ such that
$d(\ket1,S)>0$.  Assume that $d(\ket k,S)=0$ for at least one of the
basis vectors---otherwise we have the desired basis---and let $\ket
m$ be the first such basis vector in the list, i.e., $d(\ket m,S)=0$
and $d(\ket k,S)>0$ for $k<m$.  We now show that the basis can be
rotated in such a way that $d(\ket k,S)>0$ for $k\le m$. Define
\begin{equation}
\epsilon={1\over2}\min_{k<m}d(\ket k,S) \;.
\end{equation}
Let $\ket{m'}$ be a state such that $d(\ket{m'},\ket m)\equiv
d<\epsilon$ and $d(\ket{m'},S)=\delta>0$.  Then there exists a
unitary operator $\hat U$ such that $\ket{m'}=\hat U\ket m$ and
$d(\ket k,\hat U\ket k)<\epsilon$ for all $k<m$.  To see this, let
$\ket{m'}=\ket{m}e^{i\mu}\cos d+\ket{m_\perp}\sin d$, with
$\bra{m_\perp}m\rangle=0$.  We can use
$$
\hat U\equiv\ket{m'}\bra{m}+\ket{m_\perp'}\bra{m_\perp}+1-\ket m\bra
m-\ket{m_\perp}\bra{m_\perp}\;,
$$
where $\ket{m_\perp'}\equiv -\ket{m}e^{i\mu}\sin d+\ket{m_\perp}\cos
d$, for which it follows that $\cos d(\ket k,\hat U\ket k)=|\bra k
\hat U\ket k|=|1-|\bra k {m_\perp}\rangle|^2(1-\cos d)|\ge\cos d$.
Now define $\ket{k'}=\hat U\ket k$ for all $k$. Then
$d(\ket{k'},S)>\epsilon$ for $k<m$, and therefore $d(\ket{k'},S)>0$
for $k\le m$. By repeating this procedure, one arrives at a basis
with the property that each basis state is a finite distance from
$S$, as required.

\subsection{Generalized measurements}
\label{sec:genmeas}

In this subsection we investigate how the compatibility criteria
change if generalized measurements, described by POVMs, are included
in the allowed measurements.  A POVM is a collection of positive
operators $\{\hat E_b\}$ satisfying $\sum_b \hat E_b=\hat1$; the
probability assigned by party $\alpha$ to outcome $b$ is
$p_b^{(\alpha)}=\tr(\hat\rho_\alpha \hat E_b)$.

It is clear that BFM is not affected by generalizing to POVMs, since
it is phrased in terms of firm beliefs, not in terms of measurements.
For the measurement-based criteria, it is logically possible that
states that are W or W$'$ incompatible relative to ODOPs can be made
compatible by including additional measurements; indeed, the
uninformative measurement with a single outcome does make all states
W and W$'$ compatible.  Since all states are already W and W$'$
compatible under ODOPs, however, allowing POVMs makes no difference
to W and W$'$. It is also possible that states that are ES or PP
compatible relative to ODOPs can be made incompatible by including
additional measurements. It is clear, however, that density operators
with the same support satisfy ES compatibility with POVMs included
among the measurements; thus allowing POVMs makes no difference to ES
compatibility.

The only compatibility criterion that is affected by generalizing to
POVMs is PP.  We distinguish the two kinds of post-Peierls
compatibility by using PP-ODOP to denote compatibility relative to
ODOPs and PP-POVM to denote compatibility relative to POVMs.  Clearly
PP-POVM implies PP-ODOP.  To investigate PP-POVM, it is easiest to
focus on the conditions for violating PP-POVM: PP-POVM is violated if
there exists a measurement, described by a POVM $\{\hat E_b\}$, such
that at least one party assigns zero probability to every outcome
$b$.

Given any POVM, we can write the POVM elements $\hat E_b$ in terms of
their eigendecompositions, thus obtaining a finer-grained POVM
consisting of rank-one operators.  If a POVM $\{\hat E_b\}$ shows
that a set of density operators violates PP-POVM compatibility, then
the underlying rank-one POVM reveals the same incompatibility.  Thus,
in investigating PP-POVM, we can restrict attention to rank-one
POVMs. Moreover, since a rank-one POVM can be extended to an ODOP in
a higher-dimensional Hilbert space (the Neumark extension)
\cite{Peres1993a}, the question of the PP-POVM compatibility of a set
of states is equivalent to the question of whether the states are
PP-ODOP compatible when they are embedded in a Hilbert space of
arbitrary dimension.

The condition for two-party PP-ODOP compatibility,
$\hat\rho_A\hat\rho_B\ne0$, is independent of the dimension in which
the two states are embedded, so it is also equivalent to PP-POVM
compatibility \cite{Mermin2001a}. The case of two nonorthogonal pure
states, which are PP-POVM compatible, but not BFM compatible,
establishes that PP-POVM is not equivalent to BFM.

States of a two-state system (qubit) illustrate the difference
between PP relative to ODOPs and POVMs.  Since density operators of
full rank give nonzero probabilities for all measurement outcomes,
they can be added to or removed from a set of density operators
without affecting the PP compatibility of the other density operators
in the set.  For a qubit this means that we only need to consider the
situation in which the parties assign pure states
$\hat\rho_\alpha={1\over2}(\hat1+\bm{n}_\alpha\cdot\hat{\bm{\sigma}})$,
$\alpha=1,\ldots,N$, where $\bm{n}_\alpha$ is the (unit) Bloch vector
for party $\alpha$'s pure state. Two pure states are PP compatible,
relative to either ODOPs or POVMs, if and only if they are not
orthogonal.  Three or more distinct pure states in two dimensions are
PP-ODOP compatible if and only if no two of the states are
orthogonal.  PP-POVM compatibility is more complicated.  The states
are incompatible if and only if there is a POVM such that each
outcome has zero probability for at least one of the states.  Such a
POVM must consist of rank-one positive operators, each of which is
orthogonal to one of the pure states $\hat\rho_\alpha$, i.e., $\hat
E_\alpha=q_\alpha(\hat1-\bm{n}_\alpha\cdot\hat{\bm{\sigma}})$, where
$0\le q_\alpha\le1/2$.   Requiring the POVM elements to sum to
$\hat1$ implies that the $q_\alpha$'s are a normalized probability
distribution and that the Bloch vectors average to zero:
\begin{equation}
0=\sum_\alpha q_\alpha\bf{n}_\alpha\;.
\end{equation}
The result is that a set of pure states in two dimensions is PP-POVM
compatible if and only if the convex set generated by the Bloch
vectors does not contain zero or, equivalently, the convex set
generated by the states $\hat\rho_\alpha$ does not contain the
maximally mixed state $\hat1/2$.  These results for a two-state
system establish that PP-ODOP is not equivalent to PP-POVM.

We are left with the following chain of implications:
\begin{equation}
\label{eq:relquantumPOVM}
\mbox{Quantum states:}\qquad \mbox{ES} \Longrightarrow \mbox{BFM}
\Longrightarrow \mbox{PP-POVM}\Longrightarrow
\mbox{PP-ODOP}\Longrightarrow \mbox{W$'$} \Longleftrightarrow
\mbox{W}   \;.
\end{equation}
It is interesting to compare these relations to what happens to the
classical relations~(\ref{eq:relclassical}) when one generalizes to
the coarse-grained measurements that are POVMs diagonal in the common
eigenbasis of the density operators.  ES still corresponds to
concordant probabilities, BFM and PP still correspond to compatible
probabilities, but the uninformative measurement makes all
probabilities compatible under W and W$'$.  Thus we have the
following classical implications when we allow coarse-grained
measurements:
\begin{equation}
\label{eq:relclassicalPOVM}
\mbox{Classical probabilities:}\qquad \mbox{ES} \Longrightarrow
\mbox{BFM} \Longleftrightarrow \mbox{PP}\Longrightarrow
\mbox{W$'$}\Longleftrightarrow\mbox{W} \;.
\end{equation}
When we generalize to coarse-grained classical measurements, W
migrates from the strongest to the weakest compatibility condition.

\section{Three-party post-Peierls compatibility in three dimensions}
\label{sec:pp}

PP seems to be the only one of our compatibility criteria for which
there is no simple, general mathematical condition for deciding
whether a given set of density operators is compatible.  For two
parties, however, it is easy to determine whether the two density
operators are PP compatible: Mermin \cite{Mermin2001a} showed that PP
is satisfied if and only if the states are not orthogonal, i.e.,
$\tr(\hat\rho_A\hat\rho_B)\ne0$ or, equivalently,
$\hat\rho_A\hat\rho_B\ne0$.  This condition follows from the fact
that two density operators are not PP compatible if and only if there
is a measurement that can distinguish them reliably, and there is
such a measurement if and only if the two density operators are
orthogonal.  This two-party PP compatibility condition is the same
for ODOPs and POVMs, as can be shown directly \cite{Mermin2001a} or
from the fact that as an ODOP condition, it is independent of the
dimension of the Hilbert space in which the two states are embedded.
Notice that if any two parties assign PP incompatible states, then
the states of all parties are PP-ODOP and PP-POVM incompatible.

In three or more Hilbert-space dimensions, the general condition for
$N$ states to be PP compatible, relative to ODOPs or to POVMs, is
highly nontrivial.  We report results in this section for the first
interesting situation, three parties assigning states in three
Hilbert-space dimensions.  As noted above, full-rank density
operators are irrelevant to questions of PP compatibility, so we can
assume that all the density operators are either rank-one or
rank-two.  There are four cases to consider, depending on how many of
the states are pure.  We consider the three cases where one or more
of the states is mixed in Sec.~\ref{sec:mixedpure} and deal with the
case of three pure states in Sec.~\ref{sec:threepure}.

\subsection{Mixed and pure states}
\label{sec:mixedpure}

Throughout this subsection, we investigate the conditions for
constructing POVMs (or ODOPs) that show that the density operators
are PP incompatible.  Such a POVM (or ODOP) must be made up of the
operators that give zero probability for each density operator in
turn.  In doing this construction, we adopt the following
conventions.  If $\hat\rho_\alpha$ is a rank-two density operator, we
let ${\cal S}_\alpha$ denote the two-dimensional subspace that is the
support of $\hat\rho_\alpha$, with $\hat S_\alpha$ being the
projector onto ${\cal S}_\alpha$, and we let
$\hat\Pi_\alpha=|e_\alpha\rangle\langle e_\alpha|$ be the
one-dimensional projection operator that projects orthogonal to
${\cal S}_\alpha$.  A POVM element that has zero probability given
$\hat\rho_\alpha$ must have the form $\hat
E_\alpha=r_\alpha\hat\Pi_\alpha$, $0\le r_\alpha\le1$; if the POVM is
to be an ODOP, we need $r_\alpha=1$ or $r_\alpha=0$.  If
$\hat\rho_\alpha=|\psi_\alpha\rangle\langle\psi_\alpha|$ is a
rank-one density operator, we let
\begin{equation}
\hat E_\alpha=
r_{\alpha,1}\hat\Pi_{\alpha,1}+
r_{\alpha,2}\hat\Pi_{\alpha,2}
=r_{\alpha,1}|e_{\alpha,1}\rangle\langle e_{\alpha,1}|+
r_{\alpha,2}|e_{\alpha,2}\rangle\langle e_{\alpha,2}|\;,
\quad
0\le r_{\alpha,1},r_{\alpha,2}\le1\;,
\label{eq:Ealphatwo}
\end{equation}
denote the general POVM element orthogonal to $\hat\rho_\alpha$; its
eigenvectors $|e_{\alpha,1}\rangle$ and $|e_{\alpha,2}\rangle$ are
orthogonal to $|\psi_\alpha\rangle$.  A POVM element that has zero
probability given $\hat\rho_\alpha$ must have the
form~(\ref{eq:Ealphatwo}).  The operators $\hat E_{\alpha,j}=
r_{\alpha,j}\hat\Pi_{\alpha,j}$, $j=1,2$, are rank-one POVM elements
that give a fine graining of $\hat E_{\alpha}$.  If the POVM is to be
an ODOP, we must have $r_{\alpha,j}=1$ or $r_{\alpha,j}=0$.

The first (and easiest) case is that of three rank-two density
operators $\hat\rho_\alpha$.  A POVM each of whose outcomes
contradicts at least one of the parties must consist of positive
multiples of the projectors $\hat\Pi_\alpha$.  The only way three
such POVM elements can sum to $\hat1$ is if they are orthogonal
projectors (an ODOP). Thus we have that three rank-two density
operators in three dimensions are PP-POVM incompatible if and only if
the vectors orthogonal to their supports are mutually orthogonal.
Since the measurement that reveals incompatibility is an ODOP, there
is no difference between PP-POVM and PP-ODOP for this case.

A straightforward way to generalize from two parties is to say that
$N>2$ density operators are {\em pairwise PP compatible\/} if
$\hat\rho_\alpha\hat\rho_\beta\ne0$ for all pairs $\alpha,\beta$.
Though PP-POVM or PP-ODOP clearly implies pairwise PP, the converse
does not hold, as is plain from the three states
\begin{eqnarray}
\hat\rho_1&=&{1\over2}(|e_2\rangle\langle e_2|+
|e_3\rangle\langle e_3|)\;,\nonumber\\
\hat\rho_2&=&{1\over2}(|e_1\rangle\langle e_1|+
|e_3\rangle\langle e_3|)\;,\\
\hat\rho_3&=&{1\over2}(|e_1\rangle\langle e_1|+
|e_2\rangle\langle e_2|)\;,\nonumber
\label{eq:threestates}
\end{eqnarray}
which are pairwise PP compatible even though they are PP incompatible
when considered together.

The next case is that of one pure state
$\hat\rho_1=|\psi_1\rangle\langle\psi_1|$ and two rank-two density
operators, $\hat\rho_2$ and $\hat\rho_3$.  If $\hat\rho_2$ and
$\hat\rho_3$ have the same support, we are back in the situation of
pairwise PP compatibility, and the three states are PP incompatible
if and only if $|\psi_1\rangle$ is orthogonal to the common support
of $\hat\rho_2$ and $\hat\rho_3$.  Thus assume that $\hat\rho_2$ and
$\hat\rho_3$ do not have the same support.  Then $|e_2\rangle$ and
$|e_3\rangle$ span a two-dimensional subspace ${\cal R}$; denote the
projector onto ${\cal R}$ by $\hat R$.  Let $|\chi\rangle$ be the
unique (up to a phase) pure state that lies in the intersection of
${\cal S}_2$ and ${\cal S}_3$; $|\chi\rangle$ is orthogonal to ${\cal
R}$.  In addition, let $|\phi_\alpha\rangle$, $\alpha=2,3$, be the
unique (up to a phase) pure state in ${\cal S}_\alpha$ that is
orthogonal to $|\chi\rangle$; $|\phi_2\rangle$ and $|\phi_3\rangle$
lie in ${\cal R}$.

With this setup, we can turn to formulating the conditions for the
existence of a POVM that shows the density operators are
incompatible.  Such a POVM  must consist of the POVM elements $\hat
E_\alpha$ defined above.  Since only $\hat E_1$ has support outside
${\cal R}$, the only way the POVM elements can sum to $\hat1$ is to
have $|\chi\rangle$ be an eigenvector of $\hat E_1$ with eigenvalue
1, i.e., $|e_{1,1}\rangle=|\chi\rangle$ and $r_{1,1}=1$.
Consequently, $|\psi_1\rangle$ and $|e_{1,2}\rangle$ are orthogonal
vectors in ${\cal R}$, and $\hat
E_{1,2}=r_{1,2}|e_{1,2}\rangle\langle e_{1,2}|$ is a rank-one POVM
element that acts only in ${\cal R}$.  The only remaining requirement
is that $\hat R=\hat E_{1,2}+\hat E_2+\hat E_3$.  This means that we
are back to the question of constructing a POVM in two dimensions,
here the two-dimensional subspace ${\cal R}$. What we have shown is
that the PP incompatibility of the original three states is
equivalent to the PP incompatibility of the three pure states
$|\psi_1\rangle$, $|\phi_2\rangle$, and $|\phi_3\rangle$, all of
which lie in the two-dimensional subspace ${\cal R}$.

Our conclusion is the following.  The three states $|\psi_1\rangle$,
$\rho_2$, and $\rho_3$, with ${\cal S}_2\ne{\cal S}_3$, are PP-POVM
incompatible if and only if $\langle\psi_1|\chi\rangle=0$ and the
convex set generated by $|\psi_1\rangle\langle\psi_1|$,
$|\phi_2\rangle\langle\phi_2|$, and $|\phi_3\rangle\langle\phi_3|$
contains $\hat R/2$.  Similarly, the three states are PP-ODOP
incompatible if and only if $\langle\psi_1|\chi\rangle=0$ and two of
the states $|\psi_1\rangle$, $|\phi_2\rangle$, and $|\phi_3\rangle$
are orthogonal.

The third case is that of two pure states,
$\hat\rho_1=|\psi_1\rangle\langle\psi_1|$ and
$\hat\rho_2=|\psi_2\rangle\langle\psi_2|$, and one rank-two density
operator $\hat\rho_3$.  We have not been able to determine the
conditions for PP-POVM compatibility in this case, so we restrict
ourselves to PP-ODOP compatibility.  One projector in an ODOP that
reveals the incompatibility of these states must be
$\hat\Pi_3=|e_3\rangle\langle e_3|$.  The other two elements of the
ODOP, $\hat\Pi_1=|e_1\rangle\langle e_1|$ and
$\hat\Pi_2=|e_2\rangle\langle e_2|$ must operate in ${\cal S}_3$. The
states they contradict must be orthogonal to them.  Thus
$|\psi_1\rangle$ must lie in the subspace spanned by $|e_3\rangle$
and $|e_2\rangle$, and its projection onto ${\cal S}_3$, i.e., $\hat
S_3|\psi_1\rangle$, must be proportional to $|e_2\rangle$.
Similarly, $|\psi_2\rangle$ must lie in the subspace spanned by
$|e_3\rangle$ and $|e_1\rangle$, and its projection onto ${\cal
S}_3$, i.e., $\hat S_3|\psi_2\rangle$, must be proportional to
$|e_1\rangle$.  Our conclusion is that the three states are PP-ODOP
incompatible if and only $\langle\psi_1|\hat S_3|\psi_2\rangle=0$.

\subsection{Three pure states}
\label{sec:threepure}

The final case is that of three pure states in three dimensions.
Again we have not been able to prove the conditions for PP-POVM
compatibility, although we have numerical evidence that PP-POVM is
equivalent PP-ODOP for this case.  We restrict our attention in
this subsection to ODOPs.

We can assume that the states are all different, since if
two are the same, we are back in the two-party case.  Moreover, since
not being pairwise PP compatible implies not being PP compatible, the
interesting case is where the three states are pairwise PP
compatible, i.e., no pair is orthogonal.  Thus we address the
following question: under what circumstances is there an ODOP whose
outcome will definitely contradict one of three distinct,
nonorthogonal pure states?  The criterion we derive is interesting in
its own right, independent of compatibility considerations.

Let $\ket{\psi_1}$, $\ket{\psi_2}$, and $\ket{\psi_3}$ be the three
distinct, normalized, pairwise PP compatible, pure states, i.e.,
\begin{eqnarray}
&&0<|\langle\psi_1\ket{\psi_2}|<1 \;, \nonumber \\
&&0<|\langle\psi_2\ket{\psi_3}|<1 \;, \\
&&0<|\langle\psi_3\ket{\psi_1}|<1 \;. \nonumber
\end{eqnarray}
The vectors $\ket{\psi_1}$, $\ket{\psi_2}$, and $\ket{\psi_3}$
violate PP-ODOP if and only if there exist angles $\theta_k$,
$0<\theta_k<\pi/2$, $k=1,2,3$, such that
\begin{eqnarray}  \label{eq:dreibein}
a&\equiv&|\langle\psi_1\ket{\psi_2}|^2 =
(\sin\theta_1\cos\theta_2)^2 \;, \nonumber \\
b&\equiv&|\langle\psi_2\ket{\psi_3}|^2 =
(\sin\theta_2\cos\theta_3)^2 \;, \\
c&\equiv&|\langle\psi_3\ket{\psi_1}|^2 = (\sin\theta_3\cos\theta_1)^2
\;. \nonumber
\end{eqnarray}

These conditions can be seen as follows.  If the vectors violate
PP-ODOP, there exists an orthonormal basis $\{\ket1,\ket2,\ket3\}$
for the space spanned by $\ket{\psi_1}$, $\ket{\psi_2}$, and
$\ket{\psi_3}$ such that
\begin{eqnarray}  \label{eq:threevectors}
\ket{\psi_1}
  &=& e^{i\chi_1} ( \cos\theta_1\ket2 + e^{i\phi_1}\sin\theta_1\ket3 )
  \;,\nonumber \\
\ket{\psi_2}
  &=& e^{i\chi_2} ( \cos\theta_2\ket3 + e^{i\phi_2}\sin\theta_2\ket1 )
  \;,\\
\ket{\psi_3}
  &=& e^{i\chi_3} ( \cos\theta_3\ket1 + e^{i\phi_3}\sin\theta_3\ket2 ) \;,
  \nonumber
\end{eqnarray}
where $0\le\chi_k<2\pi$, $0\le\phi_k<2\pi$, $0<\theta_k<\pi/2$
($k=1,2,3$).  Taking the inner products, we see that
\begin{eqnarray} \label{eq:complex}
\langle\psi_1\ket{\psi_2}&=& e^{i(\chi_2-\chi_1)}
   e^{i\phi_1}\sin\theta_1\cos\theta_2\;, \nonumber \\
\langle\psi_2\ket{\psi_3}&=& e^{i(\chi_3-\chi_2)}
   e^{i\phi_2}\sin\theta_2\cos\theta_3\;, \\
\langle\psi_3\ket{\psi_1}&=& e^{i(\chi_1-\chi_3)}
   e^{i\phi_3}\sin\theta_3\cos\theta_1\;. \nonumber
\end{eqnarray}
The conditions~(\ref{eq:dreibein}) follow immediately.

Conversely, if the conditions (\ref{eq:dreibein}) are satisfied, then
it is clear that we can find angles $\chi_k$ and $\phi_k$ such that
the inner products $\langle\psi_i\ket{\psi_j}$ are given by
Eqs.~(\ref{eq:complex}).  Since the pairwise inner products specify
the vectors up to a unitary transformation, there exists an
orthonormal basis $\{\ket1,\ket2,\ket3\}$ such that $\ket{\psi_1}$,
$\ket{\psi_2}$, and $\ket{\psi_3}$ have the
form~(\ref{eq:threevectors}). A measurement in this basis shows that
the vectors are PP-ODOP incompatible.

\begin{figure}
\begin{center}
\includegraphics[height=3.5in]{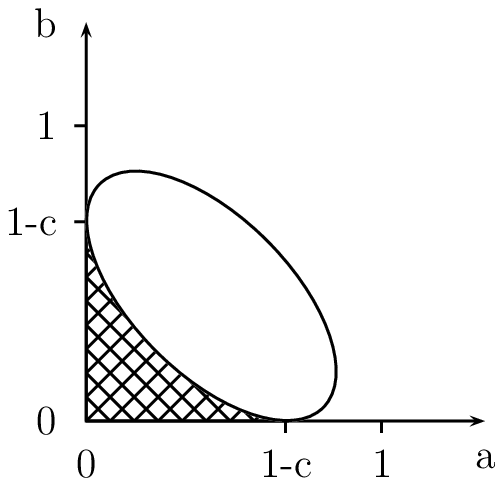}
\end{center}
\vspace{18pt}
\caption{%
Three pure states in three dimensions, $|\psi_1\rangle$,
$|\psi_2\rangle$, and $|\psi_3\rangle$, are PP-ODOP incompatible if
there exists a measurement described by three one-dimensional,
orthogonal projectors such that each outcome rules out at least one
of the three states.  The incompatibility of the states depends only
on the three squared inner products.  For a fixed value of one
squared inner product, $c=|\langle\psi_3\ket{\psi_1}|^2$, the plot
shows the region of PP-ODOP incompatibility in terms of the other two
squared inner products, $a=|\langle\psi_1\ket{\psi_2}|^2$ and
$b=|\langle\psi_2\ket{\psi_3}|^2$.  The ellipse is defined in
Eq.~(\ref{cond3}).  The region of $a$ and $b$ corresponding to
PP-ODOP incompatible states, indicated by cross-hatching, lies
between the ellipse and the axes.}
\label{fig1}
\end{figure}

To find a simpler criterion, define $x_k=\sin^2\theta_k$ for
$k=1,2,3$. The equations (\ref{eq:dreibein}) are then equivalent to
\begin{eqnarray}
 x_1(1-x_2) &=& a \;,\nonumber \\
 x_2(1-x_3) &=& b \;,\\
 x_3(1-x_1) &=& c \;.\nonumber
\end{eqnarray}
Solving these equations for, e.g., $x_2$, we obtain
\begin{eqnarray} \label{eq:quad}
x_2&=&{1-a+b-c\pm\sqrt{(1-a+b-c)^2 - 4b(1-a)(1-c)} \over 2(1-c)}
\nonumber \\
&=&{1-a+b-c\pm\sqrt{(a+b+c-1)^2 - 4abc} \over 2(1-c)} \;;
\end{eqnarray}
the expressions for $x_3$ and $x_1$ follow from cyclic permutations
of $a$, $b$ and $c$.  Equations (\ref{eq:dreibein}) are equivalent to
the existence of solutions that satisfy $0<x_k<1$ for $k=1,2,3$. The
first equality in Eq.~(\ref{eq:quad}) shows that, if there are two real
solutions, both have the same sign.  The existence of a solution
$0<x_2<1$ is thus equivalent to the following three conditions:
\begin{eqnarray}
&& 1-a+b-c > 0 \;,\label{eq:condition1} \\
&& (a+b+c-1)^2 > 4abc \;, \\
&& 1-a+b-c - \sqrt{(a+b+c-1)^2 - 4abc} < 2(1-c) \;.
\end{eqnarray}
The third of these conditions is equivalent to
\begin{equation}
-(1-b+a-c)< \sqrt{(a+b+c-1)^2 - 4abc} \;.
\end{equation}
This is implied by the condition $1-b+a-c>0$, which is a cyclic
permutation of the inequality (\ref{eq:condition1}). The full set of
conditions is therefore
\begin{eqnarray}
&& 1-a+b-c > 0 \;, \nonumber \\
&& 1-b+c-a > 0 \;, \nonumber \\
&& 1-c+a-b > 0 \;, \nonumber \\
&& (a+b+c-1)^2 > 4abc \;.
\end{eqnarray}
An equivalent form is
\begin{eqnarray}  \label{cond3}
&& |a-b| < 1-c \;, \nonumber \\
&& a+b < 1+c  \;,  \\
&& {(a+b-1)^2\over c} + {(a-b)^2\over(1-c)} > 1 \;. \nonumber
\end{eqnarray}
For fixed $c$, with $0<c<1$, it is straightforward to show that the
ellipse in the $a$-$b$ plane defined by the last inequality has the
following properties: it is centered at the point $a=b=1/2$, and its
principal axes, of length $\sqrt{c/2}$ and $\sqrt{(1-c)/2}$, form
angles of 45$^\circ$ with the $a$ and $b$ axes. The ellipse has
exactly one point of intersection with the $a$ axis at $a=1-c$ and
exactly one point of intersection with the $b$ axis at $b=1-c$.  The
ellipse and the associated region of PP-ODOP incompatibility are
shown in Fig.~\ref{fig1}.

From this it can be seen that the conditions (\ref{cond3}) are
equivalent to the following, final set of conditions,
\begin{eqnarray}
&& a+b+c < 1  \;, \nonumber \\
&& (a+b+c-1)^2 > 4abc \;,
\label{eq:3peierls}
\end{eqnarray}
which are manifestly symmetric in the three squared inner products,
$a$, $b$, and $c$.  To summarize, the three pure states are PP-ODOP
incompatible if and only if their pairwise inner products satisfy the
conditions~(\ref{eq:3peierls}).

\section{Discussion}
\label{sec:disc}

We have shown that the BFM criterion can be viewed as one member in a
hierarchy of five compatibility criteria for quantum state
assignments.  Parties whose state assignments are BFM compatible can
come to agreement about a joint state assignment without any party
having to abandon a firm prior belief.  By contrast, the four other
criteria are based on measurements.  They all have distinct roles, and
none is equivalent to BFM.

The ES criterion can be applied to a situation where all parties have
shared their available information.  They may still assign different
states, but they agree on the nullspace.  The states assigned by the
different parties all have the same support.

The PP criterion rules out the possibility of a measurement that all
parties agree will, regardless of outcome, contradict one of their
state assignments.  In other words, if the states assigned by the
parties are not PP compatible, then there exists a measurement that
will definitely reveal disagreement among the parties.  The PP
criterion is in some ways the most interesting: it puts nontrivial
constraints on the set of density operators, and it depends on
whether all generalized measurements, described by POVMs, are allowed
or the permitted measurements are restricted to ODOPs.

The W criterion (which is equivalent to W$'$) shows that there is no
strictly necessary constraint on a finite set of density operators to
be compatible.  Any such set is compatible in the sense that there
exists a measurement that allows the parties to come to agreement.
In this sense, the BFM criterion is neither sufficient nor necessary.

It turns out that there are important differences between the
classical and quantum cases.  Whereas the BFM criterion is stronger
than the PP criterion quantum-mechanically, the two are equivalent
classically.  Curiously, the criterion W, which is the weakest
quantum-mechanically, is the strongest classically, at least for
fine-grained (ODOP) measurements.

Finally, we identify {\em strong\/} Dutch-book consistency as a
necessary assumption in the derivation of the BFM criterion.  In
particular, we show that $N$ parties who violate strong Dutch-book
consistency might come to agreement about a joint state assignment
without abandoning any of their firm prior beliefs, even if their
prior state assignments are not BFM consistent.

\acknowledgments
CMC was supported in part by US Office of Naval Research Contract
No.~N00014-00-1-0578.  CMC and RS gratefully acknowledge support from
National Science Foundation Contract No.~PHY99-07949 while in
residence at the Institute for Theoretical Physics of the University
of California, Santa Barbara.  All three authors acknowledge the
hospitality of the Australian Special Research Centre for Quantum
Computer Technology at the University of Queensland, where this work
was completed.

\appendix

\section{Bayesian probabilities and the Dutch-book argument}

Bayesian probabilities are a measure of one's degree of belief in or,
equivalently, one's degree of uncertainty about the various
alternatives in a set~\cite{Bernardo1994a}.  Bayesian probabilities
receive an operational definition in decision
theory~\cite{Savage1954a}, i.e., the theory of how to decide in the
face of uncertainty.  The Bayesian approach captures naturally the
notion that probabilities represent one's beliefs about a set of
alternatives.

The simplest operational definition of Bayesian probabilities is in
terms of betting behavior, which is decision theory in a nutshell. To
formulate this definition, let $A$ be a bettor who is willing to
place a bet at odds of $(1-p)/p$ to 1 on the occurrence of an event
$E$.  These odds mean that $A$ is willing to pay in an amount
$px$---the {\it stake\/}---up front, with the promise of receiving an
amount $x$---the {\it payoff\/}---if $E$ occurs and nothing
otherwise.  To say that $A$ considers this a {\it fair\/} bet is to
say that she is willing to accept the bet at these odds no matter
what the payoff; in particular, the payoff can be either positive or
negative, meaning that $A$ is willing to accept either side of the
bet.  This situation is used to {\it define\/} probabilities: that
$A$ considers it fair to bet on $E$ at odds of $(1-p)/p$ to 1 is the
operational definition of $A$'s assigning probability $p$ to the
occurrence of event $E$.

The bookmaker who accepts the stakes and makes the payoffs is called
the {\it Dutch bookie}.  In a betting situation with $A$, he has the
freedom to set the payoffs for the various outcomes at will. $A$'s
probability assignment to the outcomes of a betting situation is
called {\it inconsistent\/} if it forces her to accept bets on which
she loses for every outcome that she deems possible.  A probability
assignment is called {\it consistent\/} (or Dutch-book consistent,
often called coherent in the literature) if it is not inconsistent in
this sense.  Remarkably, requiring consistent behavior implies that
$A$ must obey the standard probability rules in her probability
assignments: (i)~$p\ge0$, (ii)~$p(E)=1$ if $A$ believes that $E$ is
certain to occur, (iii)~$p(E\vee F)=p(E)+p(F)$ if $E$ and $F$ are
mutually exclusive, and (iv)~$p(E\wedge F)=p(E|F)p(F)$ (Bayes's
rule).  A probability assignment that violates any of these rules is
inconsistent in the above sense. This is the so-called {\it
Dutch-book argument}~\cite{DeFinettiBook,Earman1992a}, which we
review below. We stress that it does not invoke expectation values or
averages over repeated bets; a bettor who violates the probability
rules places bets that, according to her own assessment of what is
possible, will result in a sure loss in a single instance of the
betting situation.

Consider first the situation where $A$ assigns probability $p_E$ to
$E$'s occurring and probability $p_{\neg E}$ to $E$'s not occurring
(symbolized by $\neg E$).  This means that she will accept a bet on
$E$ with payoff $x_E$ (stake $p_Ex_E$) and a bet on $\neg E$ with
payoff $x_{\neg E}$ (stake $p_{\neg E}x_{\neg E}$). The net amount
$A$ receives is
\begin{equation}
G = \cases{ x_E(1-p_E) -x_{\neg E}p_{\neg E}    \!\!& if $E$
occurs,\cr -x_Ep_E + x_{\neg E}(1-p_{\neg E})  \!\!& if $E$ does not
occur.}
\end{equation}
The Dutch bookie can always choose $x_{\neg E}=0$, in which case
$A$'s gains become $G=x_E(1-p_E)$ if $E$ occurs and $G=-x_Ep_E$ if
$E$ does not occur.  To avoid all-negative gains requires that
$1-p_E$ and $p_E$ have the same sign, which implies $0\le p_E\le1$,
thus giving rule~(i).

Suppose now that $A$ believes that $E$ is certain to occur.  For the
only outcome she deems possible, her gain is $G=x_E(1-p_E) -x_{\neg
E}p_{\neg E}$.  The Dutch bookie can arrange this gain to have any
value---in particular, any negative value by choosing $x_E<0$ and
$x_{\neg E}>0$---unless $p_E=1$ and $p_{\neg E}=0$.  The result is
rule~(ii): an outcome thought certain to occur must be assigned
probability 1 (and an outcome thought certain not to occur must be
assigned probability 0).

Now consider two mutually exclusive events, $E$ and $F$, and suppose
$A$ assigns probabilities $p_E$, $p_F$, and $p_{E\vee F}$ to the
three outcomes $E$, $F$, and $E\vee F$ ($E$ or $F$).  This means $A$
will accept the following three bets: a bet on $E$ with payoff $x_E$
(stake $p_Ex_E$); a bet on $F$ with payoff $x_F$ (stake $p_Fx_F$);
and a bet on $E\vee F$ with payoff $x_{E\vee F}$ (stake $p_{E\vee
F}x_{E\vee F}$).  The net amount $A$ receives is
\begin{equation}
G  = \cases{ x_E(1-p_E) -x_Fp_F + x_{E\vee F}(1-p_{E\vee F})
    \!\!& if $E$, but not $F$ occurs,\cr
-x_Ep_E + x_F(1-p_F) + x_{E\vee F}(1-p_{E\vee F})
    \!\!& if $F$, but not $E$ occurs,\cr
-x_Ep_E -x_Fp_F -x_{E\vee F}p_{E\vee F}
    \!\!& if neither $E$ nor $F$ occurs.}
\label{eq:GEorF}
\end{equation}
We need not consider the possibility that both $E$ and $F$ occur,
since they are mutually exclusive.  The Dutch bookie can choose
payoffs $x_E$, $x_F$, and $x_{E\vee F}$ that lead to $G<0$ for all
three outcomes unless
\begin{equation}
0=\det\pmatrix{ 1-p_E &  -p_F &  1-p_{E\vee F} \cr
 -p_E & 1-p_F &  1-p_{E\vee F} \cr
 -p_E &  -p_F &   -p_{E\vee F} }
= p_E+p_F-p_{E\vee F} \;.
\end{equation}
The probability assignment is thus inconsistent unless rule~(iii) is
satisfied, i.e., $p_{E\vee F}=p_E+p_F$.

Finally, we consider two events, $E$ and $F$, which are not
necessarily exclusive.  Suppose that $A$ assigns probability $p_F$ to
the occurrence of $F$, probability $p_{E\wedge F}$ to the occurrence
of $E\wedge F$ ($E$ and $F$), and conditional probability $p_{E|F}$
to the occurrence of $E$, given that $F$ has occurred. This means $A$
will accept the following three bets: a bet on $F$ with payoff $x_F$
(stake $p_Fx_F$); a bet on $E\wedge F$ with payoff $x_{E\wedge F}$
(stake $p_{E\wedge F}x_{E\wedge F}$); and a conditional bet on $E$
given that $F$ has occurred, the payoff being $x_{E|F}$ (stake
$p_{E|F}x_{E|F}$).  If $F$ does not occur, the conditional bet is
called off, with the stake returned.  The net amount $A$ receives is
\begin{equation}
G  = \cases{ -x_Fp_F-x_{E\wedge F}p_{E\wedge F}
    \!\!& if $F$ does not occur,\cr
x_F(1-p_F)-x_{E\wedge F}p_{E\wedge F}-x_{E|F}p_{E|F}
    \!\!& if $F$, but not $E$ occurs,\cr
x_F(1-p_F)+x_{E\wedge F}(1-p_{E\wedge F})+x_{E|F}(1-p_{E|F})
    \!\!& if both $E$ and $F$ occur.}
\label{eq:GEandF}
\end{equation}
The Dutch bookie can choose payoffs $x_F$, $x_{E\wedge F}$, and
$x_{E|F}$ that lead to $G<0$ for all three outcomes unless
\begin{equation}
0=\det\pmatrix{
 -p_F   &   -p_{E\wedge F}  &   0           \cr
 1-p_F  &   -p_{E\wedge F}  &   -p_{E|F}    \cr
 1-p_F  &   1-p_{E\wedge F} &   1-p_{E|F}   }
= -p_{E|F}p_E+p_{E\wedge F} \;.
\end{equation}
Consistency thus requires that Bayes's rule be satisfied, i.e.,
$p_{E\wedge F}=p_{E|F}p_F$.

In our experience most physicists find it difficult first to accept
and then to embrace the notion that Bayesian probabilities receive
their {\it only\/} operational significance from decision theory, the
simplest realization of which is the Dutch-book argument in which
probabilities are {\it defined\/} in terms of betting odds for fair
bets.  In the Dutch-book approach, the structure of probability
theory follows solely from the requirement of consistent betting
behavior.  There is no other input to the theory.  It is worth
emphasizing, for example, that normalization is not a separate
assumption, so trivial that it requires no justification.  Rather it
is a consequence of Dutch-book consistency, specifically of
rules~(ii) and (iii), i.e., $1=p(E\vee\neg E)=p(E)+p(\neg E)$.

Surprisingly, consistency does not imply the converse of rule~(ii);
i.e., we cannot conclude from consistency alone that if $p_E=1$, then
$A$ believes that $E$ is certain to occur.  To see this, return to
Eq.~(\ref{eq:GEorF}), specializing to $p_E=1$ and $p_{\neg E}=0$ (the
latter required by normalization):
\begin{equation}
G = \cases{ 0                   \!\!& if $E$ occurs,\cr -x_E +
x_{\neg E}   \!\!& if $E$ does not occur.} \label{eq:GEE}
\end{equation}
We can get no further with consistency because the zero gain for
outcome $E$ ensures that $A$ cannot be put in a situation where all
gains are negative.

To go further, we need the notion of {\it strong consistency\/}
\cite{Shimony1955a,Kemeny1955a} (or strong Dutch-book consistency,
often called strong or strict coherence in the literature): $A$'s
probability assignment is said to be inconsistent in the strong sense
if she can be forced to accept bets on which, for outcomes she deems
possible, no gain is positive, but some gains are negative (she never
wins, but sometimes loses); a probability assignment is strongly
consistent if it is not inconsistent in the strong sense.  Since in
Eq.~(\ref{eq:GEE}) the second gain can be made negative, strong
consistency implies that $A$ must believe that $E$ is certain to
occur.  Thus strong consistency requires that $p=1$ be assigned only
to events thought certain to occur (and $p=0$ be assigned only to
events thought certain not to occur).

Dutch-book consistency requires a bettor to follow the standard
probability rules.  That following the rules is sufficient to avoid
inconsistency has been shown by Kemeny \cite{Kemeny1955a}.  Kemeny
reduces the most general betting situation to combinations of
conditional bets, as in Eq.~(\ref{eq:GEandF}), and bets on exclusive
alternatives, as in Eq.~(\ref{eq:GEorF}), and he then shows that the
expected gain for each of these kinds of bets is zero for
probabilities that satisfy the standard rules.  The expected gain for
bets on the exclusive alternatives in Eq.~(\ref{eq:GEorF}) is
\begin{eqnarray}
&\mbox{}&p_E[x_E(1-p_E)-x_Fp_F+x_{E\vee F}(1-p_{E\vee F})]
+p_F[-x_Ep_E + x_F(1-p_F) + x_{E\vee F}(1-p_{E\vee F})]\nonumber\\
&\mbox{}&\qquad+(1-p_{E\vee F})[-x_Ep_E -x_Fp_F -x_{E\vee F}p_{E\vee
F}]
\nonumber\\
&\mbox{}&\qquad =(p_{E\vee F}-p_E-p_F)[p_Ex_E+p_Fx_F-(1-p_{E\vee
F})x_{E\vee F}]\;.
\end{eqnarray}
A similar result holds for the conditional bets of
Eq.~(\ref{eq:GEandF}):
\begin{eqnarray}
&\mbox{}&(1-p_F)[-x_Fp_F-x_{E\wedge F}p_{E\wedge F}]
+(1-p_{E|F})p_F[x_F(1-p_F)-x_{E\wedge F}p_{E\wedge F}-x_{E|F}p_{E|F}]
\nonumber\\
&\mbox{}&\qquad +p_{E\wedge F}[x_F(1-p_F)+x_{E\wedge F}(1-p_{E\wedge
F})+
x_{E|F}(1-p_{E|F})]\nonumber\\
&\mbox{}&\qquad= (p_{E\wedge
F}-p_{E|F}p_F)[(1-p_F)x_F+(1-p_{E|F})x_{E|F}- p_{E\wedge F}x_{E\wedge
F}]\;.
\end{eqnarray}
Since the expected gains are zero for probabilities that satisfy the
standard rules, it is impossible to have all-negative gains (or, in
the case of strong consistency, for those outcomes the bettor deems
possible, to have gains some of which are negative with the rest
being zero).

Unlike an ordinary bookie, who tries to balance wins and losses and
makes money off the fees charged for handling the bets, a Dutch
bookie exploits inconsistencies in a bettor's behavior to win under
all circumstances (or never to lose, yet sometimes win in the case of
strong consistency).  To avoid inconsistency, a bettor simply has to
follow the rules of probability theory.  The Dutch-book argument is
not about a contest between a bettor and a Dutch bookie.  It is
wholly about the internal consistency of the way a bettor translates
beliefs into probability assignments.  The Dutch bookie is simply the
agent who exposes inconsistencies in the bettor's behavior.

In keeping with the notion that probabilities are subjective, the
Dutch-book argument does not dictate a bettor's probability
assignments, which are based on whatever the bettor believes or knows
about the situation at hand.  The only exception occurs in the case
where the bettor is certain.  Then Dutch-book consistency requires
that all her probabilities be 0 or 1.  For quantum mechanics, this
means that when a bettor is certain about the outcome of some ODOP,
she must assign the pure state corresponding to the certain outcome.
Only if the bettor is strongly consistent, however, can we conclude
that a pure-state assignment means that the better is certain about
the outcome of an ODOP that includes the pure state among its
outcomes, and this conclusion is crucial for all the compatibility
criteria developed in this paper.

\end{document}